\documentclass[preprints,article,accept,moreauthors,pdftex]{Definitions/mdpi} 
\firstpage{1} 
\pubvolume{1}
\issuenum{1}
\articlenumber{0}
\pubyear{2021}
\copyrightyear{2020}
\datereceived{} 
\dateaccepted{} 
\datepublished{} 
\hreflink{https://doi.org/} 
\pdfoutput=1

\newcommand{\lambdabar}{{\mkern0.75mu\mathchar '26\mkern -9.75mu\lambda}}

\Title{Simulation of the isotropic ultra-high energy photons flux in\,the\,solar magnetic field}

\TitleCitation{Simulation of the isotropic ultra-high energy photons flux in\,the\,solar magnetic field}

\Author{Bożena Poncyljusz$^{1*}$\orcidA{},
Tomasz Bulik$^{1}$\orcidS{},
Niraj Dhital$^{2}$\orcidT{},
Oleksandr Sushchov$^{3}$\orcidB{},
Sławomir Stuglik$^{3}$\orcidC{},
Piotr Homola$^{3}$\orcidD{},
David Alvarez-Castillo$^{3}$\orcidE{},
Marcin Piekarczyk$^{4}$\orcidF{},
Tadeusz Wibig$^{5}$\orcidG{},
Jaroslaw Stasielak$^{3}$\orcidH{},
Péter Kovács$^{6}$\orcidI{},
Katarzyna Smelcerz$^{7}$\orcidJ{},
M D Rodriguez Frias$^{8}$\orcidK{},
Michał Niedźwiecki$^{9}$\orcidL{},
Justyna Miszczyk$^{3}$\orcidM{},
Tomasz Sośnicki$^{10}$\orcidN{},
\L{}ukasz Bibrzycki$^{4}$\orcidO{},
Arman Tursunov$^{11}$\orcidP{},
Luis Del Peral$^{8}$\orcidQ{},
Krzysztof Rzecki$^{10}$\orcidR{}
}

\AuthorNames{Bo\.zena Poncyljusz, Tomasz Bulik, Niraj Dhital,  Oleksandr Sushchov, Sławomir Stuglik, Piotr Homola, David Alvarez-Castillo, Marcin Piekarczyk, Tadeusz Wibig, Jaroslaw Stasielak, Péter Kovács, Katarzyna Smelcerz, M D Rodriguez Frias, Michał Niedźwiecki, Justyna Miszczyk, Tomasz Sośnicki, \L{}ukasz Bibrzycki, Arman Tursunov, Luis Del Peral, Krzysztof Rzecki}

\AuthorCitation{Poncyljusz, B.; Bulik, T.; Dhital, N.; Sushchov, O.; Stuglik, S.; Homola, P.; Alvarez-Castillo, D.; Piekarczyk, M.; Wibig, T.; Stasielak, J.; Kovács, P.; Smelcerz, K.; Rodriguez Frias, M.; Niedźwiecki, M.; Miszczyk, J.; Sośnicki, T.; Bibrzycki, \L{}.; Tursunov, A.; Del Peral, L.; Rzecki, K.}

\address{%
$^{1}$ \quad University of Warsaw, 02-093 Warsaw, Poland; b.poncyljusz@student.uw.edu.pl\\
$^{2}$ \quad Central Department of Physics, Tribhuvan University, Kirtipur 44613, Nepal\\
$^3$ \quad Institute of Nuclear Physics Polish Academy of Sciences, Walerego Eljasza Radzikowskiego 152, 31-342, Kraków, Poland\\
$^4$ \quad Pedagogical University of Krakow, Podchorążych 2, 30-084, Kraków, Poland\\
$^5$ \quad University of Lodz, Faculty of Physics and Applied Informatics, Pomorska 149/153, 90-236, Lodz, Polska\\
$^6$ \quad Wigner Research Centre for Physics, Konkoly-Thege Miklós út 29-33., H-1121, Budapest, Hungary\\
$^7$ \quad Faculty of Computer Science and Telecommunications, Cracow University of Technology, Warszawska 24, 31-155, Kraków, Poland\\
$^8$ \quad University of Alcalá, Ctra. Madrid-Barcelona, Km. 33,7 , E-28871, Madrid, Spain\\
$^9$ \quad Department of Computer Science, Cracow University of Technology, ul. Warszawska 24, 31-155, Kraków, Poland\\
$^{10}$ \quad AGH University of Science and Technology, 30 Mickiewicza Ave., 30-059, Kraków, Poland\\
$^{11}$ \quad Institute of Physics, Silesian University in Opava, Bezručovo nám 13 , CZ-74601 , Opava, Czech Republic}

\corres{Correspondence: b.poncyljusz@student.uw.edu.pl}

\abstract{Both the lack of observation of ultra-high energy (UHE) photons and the limitations of the state-of-the-art methodology being applied for their identification motivate studies on alternative approaches to the relevant simulations and the related observational strategies. One of such new approaches is proposed in this report and it concerns new observables allowing indirect identification of UHE photons through cosmic ray phenomena composed of many spatially correlated extensive air showers or primary cosmic rays observed at one time. The study is based on simulations of interactions of UHE photons with the magnetic field of the Sun using the PRESHOWER program with some essential modifications. One of the expected results of such interactions is a generation of cosmic ray ensembles (CRE) in the form of very thin and very elongated cascades of secondary photons of energies spanning the whole cosmic ray energy spectrum. Upon entering the Earth's atmosphere, these cascades or their parts may generate uniquely characteristic walls of spatially correlated extensive air showers, and the effect is expected also in cases when primary UHE photons are not directed towards the Earth. Particle distributions in these multi-primary UHE photon footprints are expected to have thicknesses of the order of meters and elongations reaching even hundreds of millions kilometers, making them potentially observable with a global, multi-experiment approach, including re-exploring of the historical data, with the expected event rate exceeding the capabilities of even very large cosmic ray observatories. The methods described in this report allow for simulating potentially observable quantities related to CRE induced by UHE photons: densities, energy spectra and geographical orientations of particles at the top of the Earth’s atmosphere. A measurement of at least one of these quantities would be equivalent to a confirmation of the existence of UHE photons which would give an insight to fundamental physics processes at unprecedentedly high energies, far beyond the reach of man-made accelerators. On the other hand, a lack of such an observation would allow for further constraining of these fundamental processes with the physically new upper limits on UHE photon fluxes. The novel advantage of such non-observation results would lay in the purely electrodynamical character of the underlying simulations which are fully independent on extrapolations of hadronic interaction models by many orders of magnitude. Such extrapolations are necessary in the UHE photon identification methods based on the analyses of properties of individual extensive air showers presently used to determine the UHE photon upper limits.}

\keyword{cosmic rays; ultra-high energy photons; cosmic-ray ensembles}

\begin{document}


\section{Introduction}
              The nature of the cosmic rays of the highest energies is one of the unsolved problems in physics. Their sources and mechanisms of acceleration remain unknown for us, as they are the subject of ongoing research (see e.g. Ref. \cite{Symmetry} and the references therein). Therefore, the detailed analysis of physical phenomena at the highest energy regime may improve our understanding of the universe and enable us to probe modern physical theories.
       As far as we understand the interactions of subatomic particles described by the Standard Model, there should be the cut-off of the cosmic rays energy spectrum, due to the interactions of cosmic radiation components with cosmic microwave background radiation. At cosmic rays energies about 5-6$\times10^{19}$ eV the collisions of protons in the cosmic ray flux with cosmic microwave background photons may cause the production of the short-lived particle $\Delta^+$, which decays into a nucleon with the energy lower than the initial proton energy. This phenomenon is called the  Greisen–Zatsepin–Kuzmin (GZK) limit (see \cite{Greisen} and \cite{Kuzmin}). In turn, heavier nuclei of cosmic radiation at energies about 10$^{20}$ eV may undergo photodisintegration and in this way also lower their energy. Therefore, the maximum distance from the source for the possible observation of the highest energy cosmic rays should be limited. \\
       Recent experimental data \cite{PDG} indicate that the cosmic rays energy spectrum is suppressed, but it is impossible to determine whether the observations are consistent with the mentioned theory. On the other hand, the nature of detected cosmic radiation particles with energies exceeding 10$^{20}$ eV is unknown, as there are no potential sources of such particles in the Earth vicinity. Thus, the investigation of ultra-high energy cosmic rays is significant.
    
       The problem of the highest energy cosmic rays gave rise to new theoretical models and predictions (see e.g. Ref. \cite{Symmetry} and the references therein), that can be justified or falsified only by detailed analysis of processes at the ultra-high energy regime. One of these predictions is the existence of the Super Heavy Dark Matter, which decay or annihilation may lead to the production of ultra-high energy cosmic ray flux composed mainly of photons (see Ref. \cite{SHDM_Kuzmin}, \cite{SHDM_Birkel}, \cite{SHDM_Frampton} and \cite{SHDM_Blasi}). Hence, the study of ultra-high energy photons can be considered as an indirect search for heavy dark matter particles. Similarly, the rest of the new physical theories predict ultra-high energy (UHE) photon creation. However, the expected UHE photons flux is much higher than the observed one.
       A possible explanation for lowering the UHE photons flux is the cosmic ray ensembles (CRE) phenomenon. Providing that a UHE photon is cascading already in space, then a group of secondary cosmic rays might reach the top of the atmosphere instead of an individual particle. Thus, instead of only one extensive air shower (EAS) created in the atmosphere, a group of (EAS) is generated. This correlation may involve arrival time relations and/or spatial distributions to provide a characteristic signature, although a detection of this phenomenon might require an extended detector array, even about the size of the Earth, as proposed by the Cosmic-Ray Extremely Distributed Observatory (CREDO)\cite{Symmetry}. The project aims at analysing data from professional observatories, educational cosmic ray facilities, and also from the smartphones with the CREDO Detector  \cite{detector, detector_paper} mobile application, where a CMOS camera is used as a particle detector. For proper collected data interpretation dedicated to identification of CRE, the process of UHE photon cascading, formation of correlated air showers, and detector response need to be correctly modelled. This is the reason why, among other researches, the CREDO project probes the scenario of UHE photon cascading in the solar magnetic field \cite{JCAP}. The specific methods dedicated to utilizing this phenomenon for planning novel indirect UHE photon searches are presented in this article.

       The previous studies of the CRE effect \cite{Symmetry, JCAP} showed that it is possible to observe some photons from cascades generated by the primary UHE photons also in cases where they are not aimed at the Earth. Here, we use the PRESHOWER program \cite{preshower2005,PRESHOWER} with modifications to include such cases in the simulations for the first time, and deliver complete distributions of CRE photons at the top of the atmosphere which point to relevant observational strategies based on the properties of these distributions.
       \section{Model of ultra-high energy photon interaction}
       We modelled the Sun's vicinity by a surrounding sphere with radius $6\ \mathrm{R_{\odot}}$. It was also assumed that UHE photons flux reaching the start sphere is isotropic. However, in the case of the UHE photon direction, only photons that come from outside of the sphere are considered to avoid the case duplication. Limits on the UHE photons are applied in accordance with \cite{Auger}. Then, the cascading of UHE photon in the solar magnetic field is considered.\\
       Physical processes that are responsible for cascading the UHE photon in space are magnetic pair production, the bending of the charged particle trajectory, and synchrotron photon emission. In the considered case, a primary UHE photon may convert to a positron and electron in a sufficiently strong magnetic field. Then, trajectories of the created pair are bending, due to the presence of a magnetic field. Hence, the positron and the electron emit synchrotron photons, that may be observable as photons cascade on the top of the atmosphere. The following mathematical description of the UHE photon cascading is based on Ref. \cite{preshower2005} and the references therein.
       The mathematical formulations of quantum mechanics of magnetic pair production used in this work were taken from \cite{Erber} and \cite{Daughtery}. On the basis of \cite{Erber}, the number of pairs created by $n_{photons}$ photons in transverse magnetic field $B$ is described by the following expression
        \begin{equation}
            n_{pairs}=n_{photons}\left[1-\exp({-\alpha(\chi)dl})\right],
            \label{conv}
        \end{equation}
        where $dl$ is photon path length and $\alpha(\chi)$ is the attenuation coefficient, that depends on parameter
        $$\chi=\frac{1}{2}\frac{h\nu}{m_ec^2}\frac{B}{B_{cr}},$$
        where $h\nu$ - photon energy, $m_e$ - electron mass and $B_{cr}=4.414\times10^{13}$ G is the critical magnetic field strength.The attenuation coefficient depends on the probability transition for magnetic pair production. Using the first-order perturbation method and assuming that $B\ll B_{cr}$, the approximated formula for $\alpha(\chi)$ may be expressed as
        $$\alpha(\chi)=\frac{1}{2}\frac{\alpha_{QED}}{\lambdabar_c}\frac{B}{B_{cr}}T(\chi),$$
        where $\alpha_{QED}$ is the fine-structure constant, $\lambdabar_c$ is reduced Compton wavelength and auxiliary function $T(\chi)$ is approximated by
        \begin{equation}
            T(\chi)\approx \frac{0.16}{\chi}K^2_{1/3}\left(\frac{2}{3\chi}\right)
            \label{eq:T-function}
        \end{equation}.
        Whereas, the $K_{1/3}\left(\frac{2}{3\chi}\right)$ in the expression (\ref{eq:T-function}) is modified Bessel function.\\
        According to equation (\ref{conv}), the probability of conversion within infinitesimal distance $dl$ is
        \begin{equation}
            p_{conv}=1-\exp({-\alpha(\chi)dl})\approx\alpha(\chi)dl
            \label{eq:prob_conv}
        \end{equation}
        and for larger distance $L$ it should be calculated as
        $$P_{conv}=1-\exp{\left[-\int^L_0\alpha(\chi)dl\right]}.$$
        The conversion rate is negligible unless the condition $\chi=\frac{1}{2}\frac{h\nu}{m_ec^2}\frac{B}{B_{cr}}\approx0.1$ is satisfied. Thus, magnetic pair production has not been observed so far.\\
        The parameter $\chi$ also determines the shape of the pair-member energy distribution, which is given by the expression
        \begin{equation}
            \frac{dn}{d\epsilon} \approx \frac{\alpha_{QED}}{\lambdabar_c}\frac{B}{B_{cr}}\frac{\sqrt{3}}{9\pi\chi}\frac{[2+\epsilon(1-\epsilon)]}{\epsilon(1-\epsilon)}K_{2/3}\left[\frac{1}{3\chi\epsilon(1-\epsilon)}\right],
            \label{eq: frac_energ}
        \end{equation}
        where $\epsilon$ is a fractional pair-member energy. The derivation of this formula is presented in detail in \cite{Daughtery}.
       The radial acceleration of electron and positron in the solar magnetic field is calculated using
        $$\frac{d\mathbf{\hat{v}}(t)}{dt}=\frac{qc^2}{E}\mathbf{\hat{v}}\times\mathbf{B},$$
        where $\mathbf{\hat{v}}$ is a velocity versor, $q$ is particle charge, $E$ is particle kinetic energy and $\mathbf{B}$ is an external magnetic field. For short time interval $\Delta t$, altered particle direction $\mathbf{\hat{v}}(t+\Delta t)$ may be approximated with Taylor series expansion
        $$\mathbf{\hat{v}}(t+\Delta t) \approx \mathbf{\hat{v}}(t) + \frac{d\mathbf{\hat{v}}(t)}{dt}\Delta t$$
        and in this case
        \begin{equation}
           \mathbf{\hat{v}}(t+\Delta t) \approx \mathbf{\hat{v}}(t) + \frac{qc^2}{E}(\mathbf{\hat{v}}\times\mathbf{B})\Delta t\ .
           \label{traj}
        \end{equation}
        In the PRESHOWER simulator an appropriate time interval $\Delta t$ is chosen according to the particle energy and the magnetic field variability. During the first half of the $\Delta t$ period, the particle propagates in the previous direction. Then the particle direction is recalculated in accordance with (\ref{traj}) and in the second half of the $\Delta t$ particle is moving in an altered direction. In the end, the particle direction is recalculated once again.\\
        The energy distribution of emitted synchrotron radiation for electron at ultra-relativistic regime, according to \cite{Sokolov}, is expressed by
        $$f(y)=\frac{9\sqrt{3}}{8\pi}\frac{y}{(1+\xi y)^3}\left[\int^{\infty}_y K_{5/3}(z)dz + \frac{(\xi y)^2}{1+\xi y}K_{2/3}(y)\right],$$
        with parametrization $\xi=\frac{3}{2}\frac{B_{\perp}}{B_{cr}}\frac{E}{m_ec^2}$, where $E$ is electron/positron energy, and
        $$y(h\nu)=\frac{h\nu}{\xi(E-h\nu)},$$
        where $h\nu$ is emitted photon energy.
        In turn, energy emitted by electron or positron within path length $dl$ is equal to
        $$W_{dl}=P\frac{dl}{c}=\frac{2}{3} \frac{r^2_0c}{(m_ec^2)^2}E^2|\mathbf{B}|^2\frac{dl}{c},$$
        where $P$ is radiation power and $r_0$ is classical electron radius.
        Therefore, one can obtain that the probability of emission of synchrotron photon within infinitesimal path length $dl$ is
        \begin{equation}
            P_{brem}(B_{\perp},E,h\nu,dl) = dl\int^E_0 I(B_{\perp},E,h\nu)\frac{d(h\nu)}{h\nu},
            \label{eq: brem}
        \end{equation}
        where
        $$I(B_{\perp},E,h\nu)=\frac{h\nu dN}{d(h\nu)dl} = f(y)\frac{W_{dl}}{dl}\frac{dy}{d(h\nu)}$$ and $dN$ - the number of photons with energy $[h\nu,\ h\nu+d(h\nu)]$, emitted within the path length $dl$.
        According to \cite{Wiedemann}, at ultra-relativistic regime, synchrotron photons are emitted in half opening angle
        \begin{equation}
            \theta\approx\frac{1}{\gamma}
            \label{eq: direction}
        \end{equation}
        about charge particle direction, where $\gamma$ is the Lorentz factor. The azimuthal angle of emitted photon direction is random, consistent with uniform distribution. Therefore, at the ultra-relativistic regime the direction of the emitted synchrotron photon is nearly tangential to the electron or positron trajectory. In this way, synchrotron photons may reproduce the projection of a pair-member trajectory onto the atmosphere plane. For that reason the spatial distribution of secondary photons on the top of the atmosphere is expected to be a very thin curve.\\
        Due to natural variations and complexity of the solar magnetic field, it is impossible to apply a satisfying, well approximated model in the PRESHOWER simulator. Thus, the solar magnetic field is modelled by the simple dipole approximation. The magnetic dipole moment used in simulations equals $6.87\times10^{32}\ \mathrm{G\ cm^3}$.
 
\section{Materials and Methods}
    In order to model the UHE photon cascading in the solar magnetic field, the PRESHOWER program is used in this work. In this program, interactions are modelled by Monte Carlo simulations. However, the program had to be modified to enable the time cumulative distribution of secondary photons on the top of the atmosphere. The detailed description of the PRESHOWER program is presented in \cite{PRESHOWER}. Unfortunately, the article refers to the simulator of the UHE photon interaction with the geomagnetic field, so the simulator version applied in this work is different. However, this section and the indicated article provide sufficient knowledge to understand research methods.\\
    The PRESHOWER simulator generates sample data for determined initial conditions. Simulation parameters that could be changed were the primary photon energy, the impact parameter, and the heliocentric latitude of the primary photon trajectory. Therefore, only two coordinates of the simulation start point could be adjusted. What is more, the primary photon direction was completely determined after the choice of parameters. For these reasons, it was necessary to update the PRESHOWER simulator, with the purpose of applying the more realistic model. The introduced modifications and program functioning are described in the following sections.\\

\begin{figure}
        \includegraphics[width=10.5 cm]{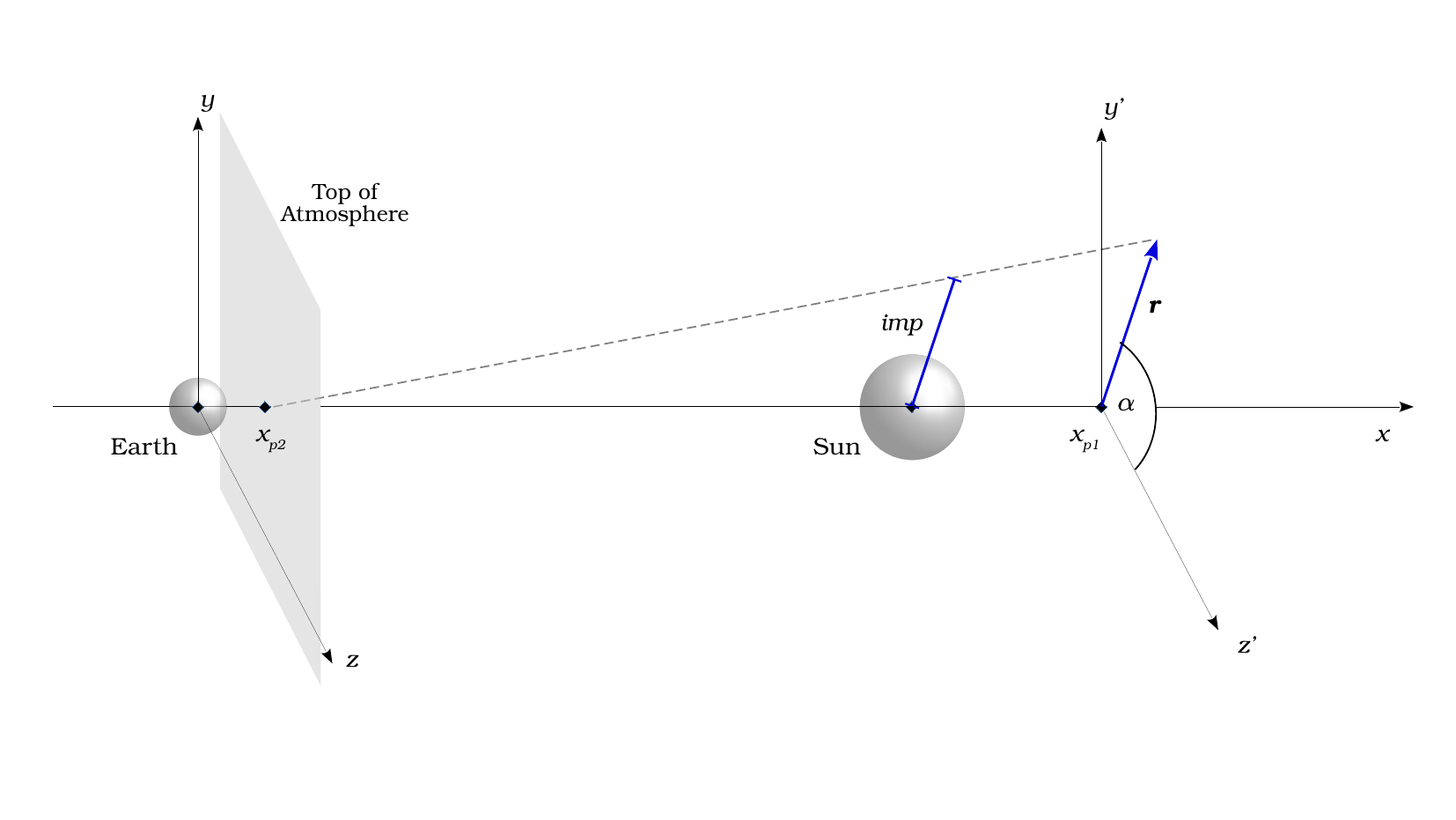}
        \caption{The coordinate system used in the previous version of the PRESHOWER simulator. The dashed line shows the primary photon trajectory, $x_{p1}$ - the fixed x coordinate of the simulation start point, the distance between $x_{p1}$ and the Sun's centre is $10\ R_{\odot}$. The $imp$ (blue line) is the impact parameter (the input simulation parameter) and the blue vector $\textbf{r}$, which lies in the $y'z'$ plane, points at the simulation start point. The angle $\alpha$ between the $\textbf{r}$ and the $z$ axis is the heliocentric latitude (the simulation input parameter), while the vector length $|\textbf{r}|=r\_at\_beginning$ is a parameter calculated during the simulation. The $x_{p2}$ point determines the top of the atmosphere plane $x=x_{p2}$ and it is the land point of the primary photon.}
        \label{fig:uklad_pop}
    \end{figure}

    The coordinate system used in the simulator is shown in Figure~\ref{fig:uklad_pop}. The applied coordinate system is similar to the heliocentric ecliptic coordinate system, but in contrast to this system, the Earth motion is not simulated. Therefore, the $x$ axis is determined by the centre of the Earth and the centre of the Sun. The Earth centre is also the coordinate system origin. The $xy$ plane contains the ecliptic. In fact, the simulator neglects the Earth and its atmosphere curvature. The atmosphere flat is modelled by a plane with equation $x=112.83$ km. On the scheme, simulation parameters - the impact parameter and the heliocentric latitude of the primary photon trajectory are indicated. The impact parameter in this model is the distance between the primary photon trajectory and the Sun centre. Whereas, the heliocentric latitude is defined as the angular distance between the ecliptic and a simulation start point in the heliocentric coordinate system. In this way, the simulation start point is completely determined, because the $x$ coordinate of the start point is a constant value for all simulations equals $156.9\times10^6$~km, which is implemented in the code and cannot be adjusted.
    Moreover, the primary photon is always aimed at the centre of the atmosphere plane. Therefore, it is assigned after a simulation parameters choice. In this case, the heliocentric latitude remains unchangeable throughout the primary photon propagation.
    
    For proper functioning of the program, many functions with different destinations need to be applied, among others, the CERN routine DBSKA for calculating the modified Bessel functions. The most important function of the simulator performs a loop on particles and simulates physical processes involving them and propagates them. The loop includes only the primary UHE photon and possibly the primary pair, because secondary photons' directions are not altered and they do not need to be propagated step-wise. Thus, their position at the top of the atmosphere is calculated just after their generation and potential secondary photons cascading is neglected.
    \\
    The function calculates a simulation start point for given input parameters. Then, it propagates UHE photon and simulates its conversion gradually. In each step, the probability for conversion is computed according to the formula (\ref{eq:prob_conv}).
    
    For a stochastic sampling of the probability for the process occurrence, the long period pseudo-random number generator is used. Generated numbers are consistent with the uniform distribution. Therefore, thanks to imposing the condition that the process occurs if the generated number is less than the calculated probability, the simulation reflects the probability distribution for the considered process. When the electron-positron pair is created, the fractional energy is simulated similarly to the conversion probability, with respect to its probability distribution (\ref{eq: frac_energ}).
    \\
    Then, the electron and positron are propagated step-wise according to the formula (\ref{traj}). The distance $\Delta s$ chosen in each step, which refers to the time interval $\Delta t$, depends on the magnetic field strength for the particle location in space and the current particle energy. Secondary photons emission is simulated by analogy with the primary photon conversion. The probability for this process is computed on the basis of the formula (\ref{eq: brem}). And the approximation (\ref{eq: direction}) is adopted as the angle of synchrotron photon deflection from the electron or positron trajectory. In turn, the angle of photon direction in the plane horizontal to the emitting particle trajectory is sampled from the uniform distribution. The simulation ends when the primary photon or created particles reach the top of the atmosphere. All obtained particle data is stored in a matrix, which has a fixed size. Therefore, the maximum number of particles generated in the simulation is limited. 
    
    On the basis of the simulation results, the spatial distribution of secondary photons at the top of the atmosphere may be analysed in detail, in the case of the UHE photon aimed at the top of the atmosphere. The influence of simulation input parameters on the distribution has been the subject of the previous study \cite{JCAP}.\\
    
\begin{figure}
       \includegraphics[width=10.5 cm]{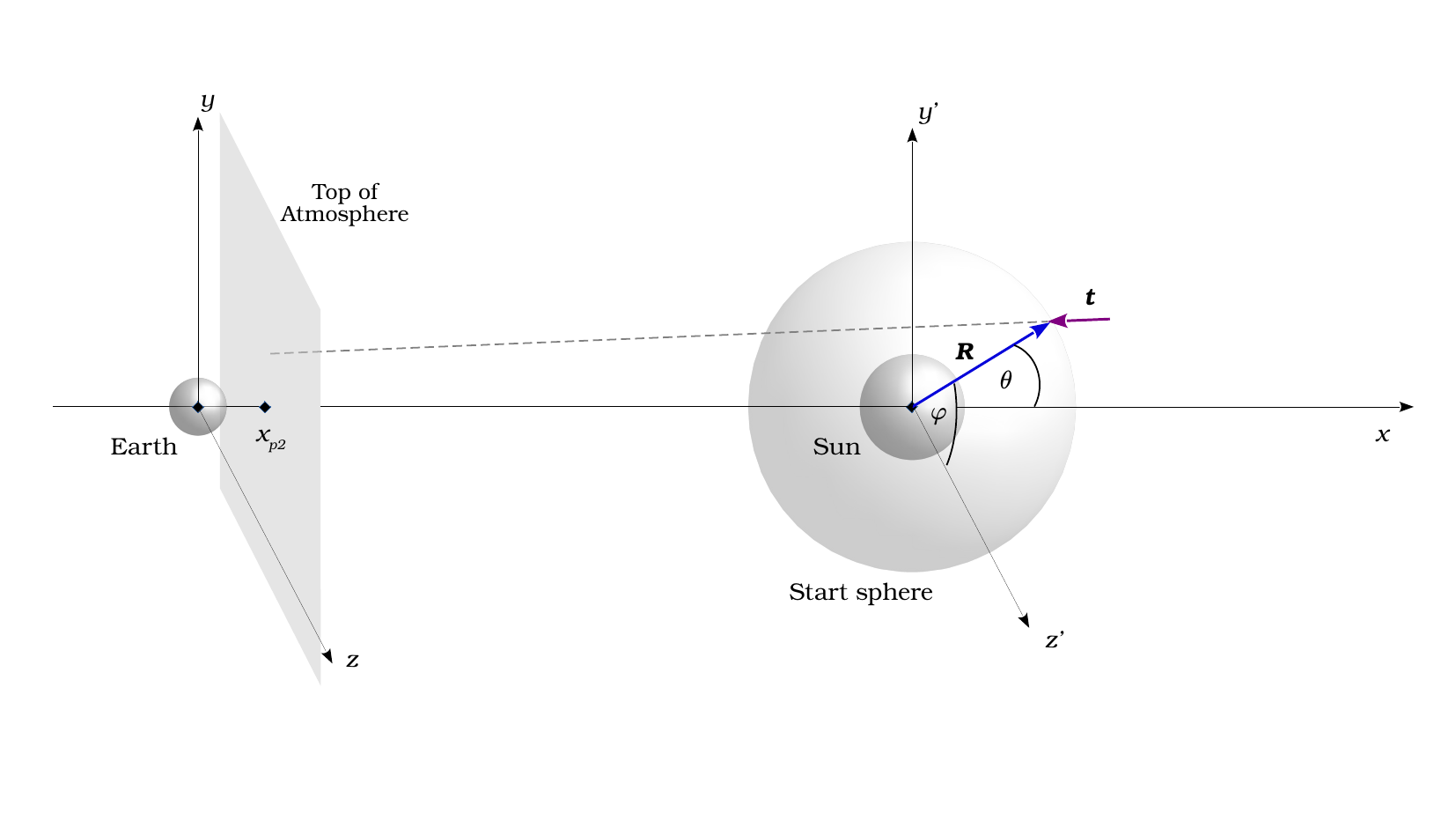}
       \caption{The coordinate system used in the modified version of the PRESHOWER simulator. The dashed line shows the primary photon trajectory. The purple vector $\textbf{t}$ is the primary photon direction, which is randomly generated. The blue vector $\textbf{R}$ points out the simulation start point, its length $R$ is the simulation input parameter, whereas the polar angle $\theta$ between the vector $\textbf{R}$ and the $x$ axis and the azimuthal angle $\varphi$ in the $y'z'$ plane are randomly generated. The $x_{p2}$ point determines the top of the atmosphere plane $x=x_{p2}$, but it is no longer the primary photon land point.}
       \label{fig:uklad_teraz}
   \end{figure}

    Previous results indicated that it is justified to consider a more realistic case, in which UHE photons reaching the Sun's vicinity from every direction are modelled. Furthermore, it was assumed that the UHE photons flux is isotropic. Nevertheless, this approach required substantial changes in the program's functioning.\\
   The most crucial modification concerned the arbitrary direction of the UHE photon, which implies the need for new simulation parameters and a new method of calculating the simulation start point. On the basis of assumptions made about the UHE photons flux, it was found that the most suitable for simulations is the heliocentric coordinate system. Figure~\ref{fig:uklad_teraz} illustrates the modified coordinate system. It is worth mentioning that the inclination angle $\theta$ in the coordinate system is defined as an angle between the $x$ axis and the position vector.
   In the new version of the PRESHOWER, simulation start points should be isotropically distributed at the sphere surface. Therefore, the radius of this start sphere was chosen as the simulation parameter. Whereas, the position on the start sphere is generated in such a way that every solid angle is equally probable. This condition may be described as
   \begin{equation}
       \frac{dN}{d\Omega}=\mathrm{const}\equiv C ,
   \end{equation}
   where $dN$ is the number of UHE photons at the start sphere in the solid angle $d\Omega$. Therefore, the probability of generating the primary photon at the position with the azimuthal angle from interval $[\varphi, \varphi+d\varphi]$ should be equal
   \begin{equation*}
      P_{\varphi}=\frac{1}{N}\int_{\Omega_0}\frac{dN}{d\Omega}d\Omega=\frac{1}{N}\int^{\pi}_0 \sin\theta d\theta \int^{\varphi+d\varphi}_{\varphi}Cd\varphi=\frac{C}{N}d\varphi ,
   \end{equation*}
   where $N$ is the total number of UHE photons reaching the start sphere. The probability density function of the azimuthal angles is constant
   \begin{equation}
      f(\varphi)=\frac{dP_{\varphi}}{d\varphi}=\frac{C}{N}=\mathrm{const}
      \label{eq:phi_prob_dens}
   \end{equation}
   and the variable is described by the uniform distribution.
   Similarly, one may obtain that
   \begin{equation}
      f(\cos\theta)=\mathrm{const}
      \label{eq:theta_prob_dens}
   \end{equation}
   and $\cos\theta$ is uniformly distributed. Therefore, in the applied function $\varphi$ and $\cos\theta$ are randomly generated from uniform distribution and then the simulation start point is calculated using the standard transformation between spherical coordinates and Cartesian coordinates. In the transformation, the redefinitions of the $\theta$ angle and the Sun's position are included. 
   The isotropic primary photon direction is modelled analogically. Nevertheless, the transformations from $(\tilde{\theta},\tilde{\varphi})$ coordinates to coordinates used in the simulation are more complex. Due to the fact that $(\tilde{\theta},\tilde{\varphi})$ are defined in the local coordinate system related to the position on the start sphere. Generated angles are firstly transformed to local Cartesian system
  \begin{gather*}
   \hat{e}_{\tilde{x}} = -\sin\theta\cos\varphi,\quad
   \hat{e}_{\tilde{y}} = -\sin\theta\sin\varphi,\quad
   \hat{e}_{\tilde{z}} = -\cos\theta.
   \label{eq: dir1}
  \end{gather*}
  The minus sign results from considering primary photons coming from the outside of the start sphere. The local versors may be identified as spherical versors 
  \begin{equation}
      \hat{e}_{\tilde{z}}=\hat{e}_r ,\quad
      \hat{e}_{\tilde{x}}=\hat{e}_{\theta} ,\quad
      \hat{e}_{\tilde{y}}=\hat{e}_{\phi}
  \end{equation}
   related to the position on the start sphere $(\theta_S,\varphi_S)$. Therefore, they are expressed in coordinates used in the simulator (Fig. \ref{fig:uklad_pop}) using formulas
  \begin{gather*}
   \hat{e}_x = \hat{e}_{\tilde{z}}\cos\theta_S-\hat{e}_{\tilde{x}}\sin\theta_S\\
   \hat{e}_y = \hat{e}_{\tilde{z}}\sin\theta_S\cos\varphi_S-\hat{e}_{\tilde{y}}\sin\varphi_S+\hat{e}_{\tilde{x}}\cos\theta_S\cos\varphi_S\\
   \hat{e}_z = \hat{e}_{\tilde{y}}\cos\varphi_S+\hat{e}_{\tilde{x}}\cos\theta_S\sin\varphi_S+\hat{e}_{\tilde{z}}\sin\theta_S\sin\varphi_S.
   \label{eq: dir2}
  \end{gather*}
  Additionally, the drawn value $\cos\tilde{\theta}$ belongs to the interval $[0,\ 1]$, since we consider only the range of polar angles $[0,\ \pi/2]$, which responds to directions beneath the sphere surface.
 
  The generalisation of the program for an arbitrary primary photon direction required also changes in the simulation end condition. The previous condition may be effectively applied only to primary photons aimed at the atmosphere plane. Thus, assuming the plane is infinite, the previous condition is placed on photons with the negative component of the direction versor. In other cases, the simulation end condition is the distance to the Sun's centre larger than 7~$R_{\odot}$. What is more, the propagation of the primary photon, electron, or positron is aborted in case the particle lands on the Sun. In turn, secondary photons aimed at the Sun are not even included in the particle table. In the previous version of the program, these additional conditions were not needed, because of the specific determination of the primary photon direction.
  This approach enables considering also the potential special cases of the UHE photon cascading. For instance, it takes into account the primary photons not aimed at the Earth, that may generate observable secondary photons after conversion, thanks to the change of electron or positron direction in the magnetic field.




\section{Results}
    Introduced amendments allowed to simulate the cumulative distribution of secondary photons at the top of the atmosphere. However, the new version of the simulator required to be verified in detail beforehand. Therefore, both the validation of the simulator and the preliminary analysis of results are described in the following sections.
\subsection{Verification of the PRESHOWER functioning}
\begin{figure}
           \includegraphics[width=10.5 cm]{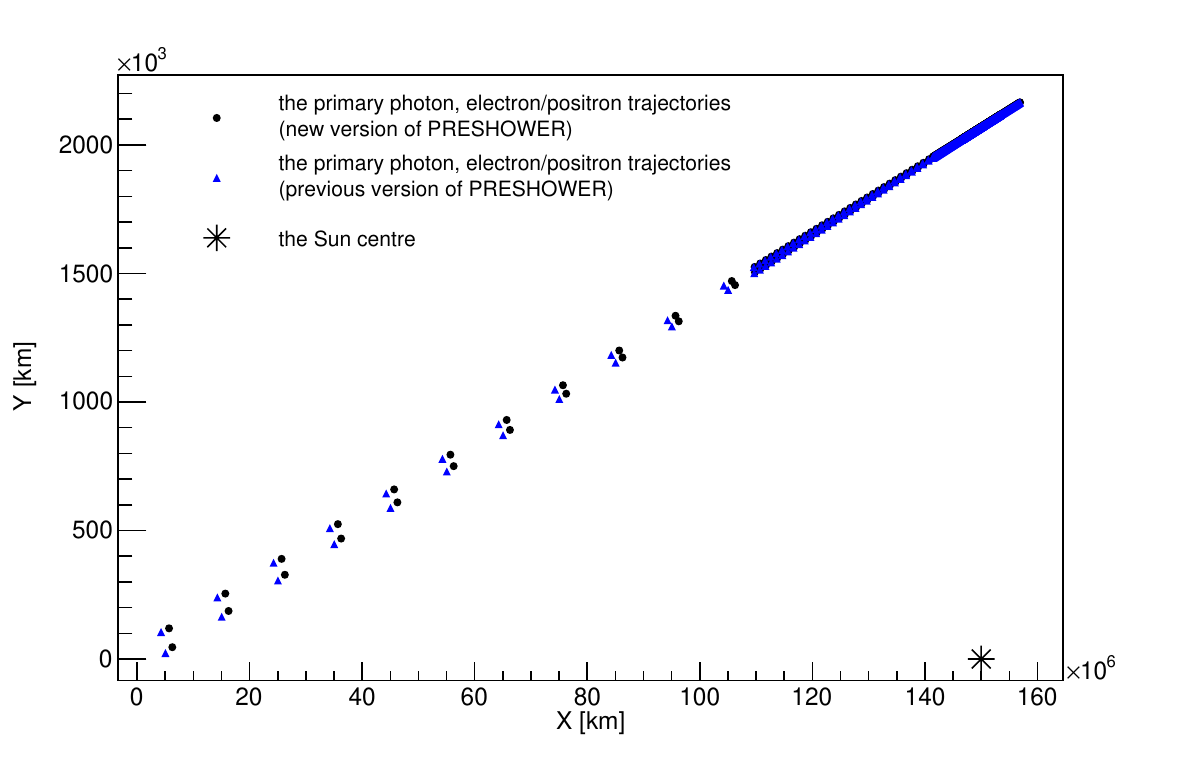}
           \caption{The comparison of particles (the primary photon and the positron-electron pair) propagation in the previous and modified version of PRESHOWER, for the exemplary simulation with the primary photon energy 100 EeV.}
           \label{fig: example_sim}
       \end{figure}

\begin{figure}
        \includegraphics[width=10.5 cm]{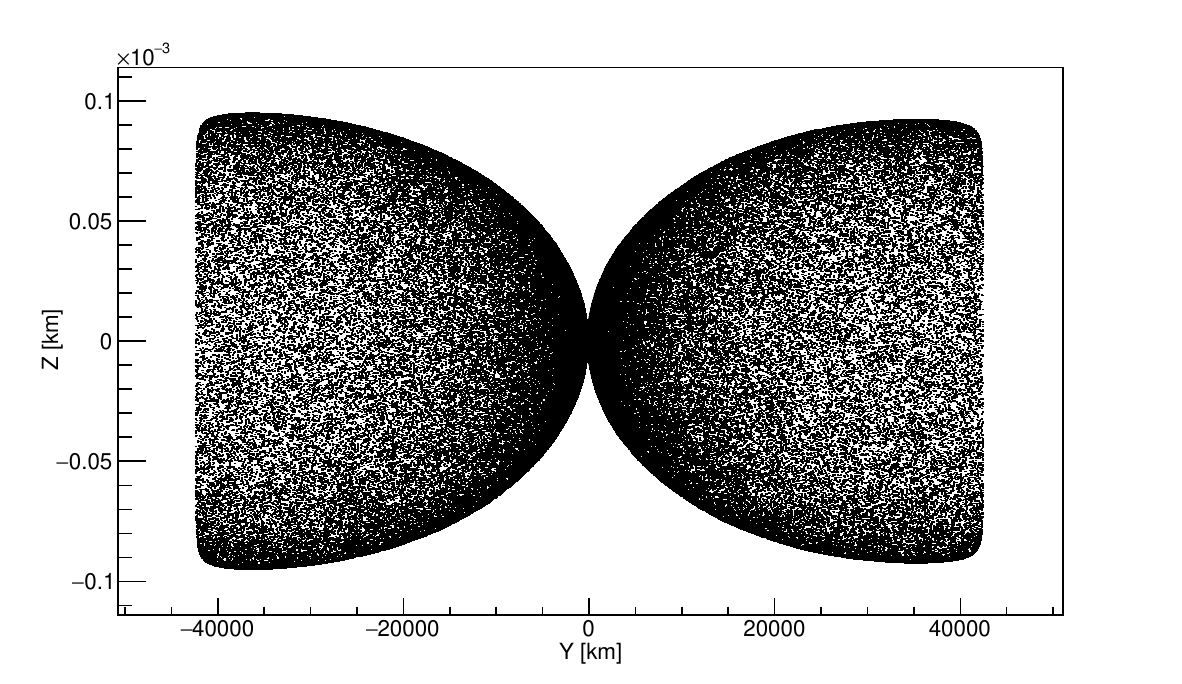}
        \caption{The spatial distribution of secondary photons at the top of the atmosphere produced in the exemplary simulation \ref{fig: example_sim}. Note the difference in the scales along the $y$ and $z$ axes.}
        \label{fig:exa_sim_sps}
    \end{figure}

    It was significant to analyse simulation results firstly with respect to the correctness. This verification was performed both for appended functions and for the simulation as a whole. In the case of added functions, we visualised the way that examples of results model the phenomena. For this purpose, we analysed the distribution of generated solid angles and then visualised calculated positions at the sphere and directions. On the basis of 100 000 generated $\cos\theta$ and $\varphi$ values, it was found that the distribution of solid angles is flat. Similarly, the visualisation of position and direction distributions were in line with our expectations.
    With the purpose of verifying the whole program and the interoperability of its functions, it was necessary to compare results from the present version of the PRESHOWER with those from the previous version. However, this comparison should concern an identical physical event. Therefore, the simulation start point, the primary photon direction, and energy must be the same in both used simulator versions. We considered the case when the primary photon with energy 100 EeV aimed at the Earth propagates in the Sun's vicinity with the impact parameter 3 $R_{\odot}$ and the heliocentric latitude 0$^{\circ}$. This responds to following parameters in the current PRESHOWER version: the simulation start point $(R, \theta_S,\phi_S)=(10.44\ R_{\odot},\ 0.29,\ 0)$, where $R$ is the start sphere radius, and the primary photon direction $(\tilde{\theta},\tilde{\phi})=(0.28,\ \pi)$.
    In this simple case the primary photon, electron, and positron are propagating only in the $xy$ plane. Their movement modelled by both simulations is illustrated in Figure~\ref{fig: example_sim}. The graph shows particle positions for each thousandth step of the simulation.
    It was found that the particles’ propagation is similar in both cases. Minor differences such as distances between the next steps are caused by a different conversion point. This, in turn, results from the stochastic nature of physical processes.
    Furthermore, the distribution of secondary photons at the top of the atmosphere is presented in Figure~\ref{fig:exa_sim_sps}. The generated distribution has a line-like shape and a size $8.48\times10^4$ km, which is a reasonable outcome. In accordance with previous results \cite{JCAP}, the distribution size should be approximately $10^5$ km.
    Therefore, obtained results are correct and it was concluded that the simulator properly models the UHE photon cascading.
    \subsection{Optimization of the simulation}
    The inclusion of more cases in the simulation, unfortunately, extended the computation time. It was found that the simulator needs to be optimised. Due to this fact, we reduced the cases included by the program and adjusted its parameters.
    \subsubsection{Reducing considered cases}
    First simulations showed that only a few of them gave rise to observable effects on the Earth. Thus, most of the calculation time was used for cases that are not the subject of this study. With the purpose of the effective disposal of computer resources, we resigned from modelling the special cases of the UHE photon cascading. Additional conditions were placed on the primary photon direction and its land point at the top of the atmosphere. Only primary photons moving towards the negative $x$ axis are simulated. In turn, the primary photon land point is usually related to the centre of secondary photons distribution. In this way, limiting its distance from the $x$ axis (Figure~\ref{fig:uklad_teraz}) increases the probability of the observation on the Earth. The limit on the distance between the point at the top of the Earth’s atmosphere where the UHE photon would have landed if no conversion had occurred, and the $x$ axis connecting Earth's and Sun's centres is hereinafter referred to as the land point limit. Conditions are checked at the beginning of the program and further procedures are performed only for primary photons that fulfil limits.
    What is more, the number of secondary photons saved to the outfiles was limited to particles that are not further than 6500 km from the $x$ axis at the top of the atmosphere.
    \subsubsection{Adjustment of simulation parameters}
\begin{figure}
        \includegraphics[width=10.5 cm]{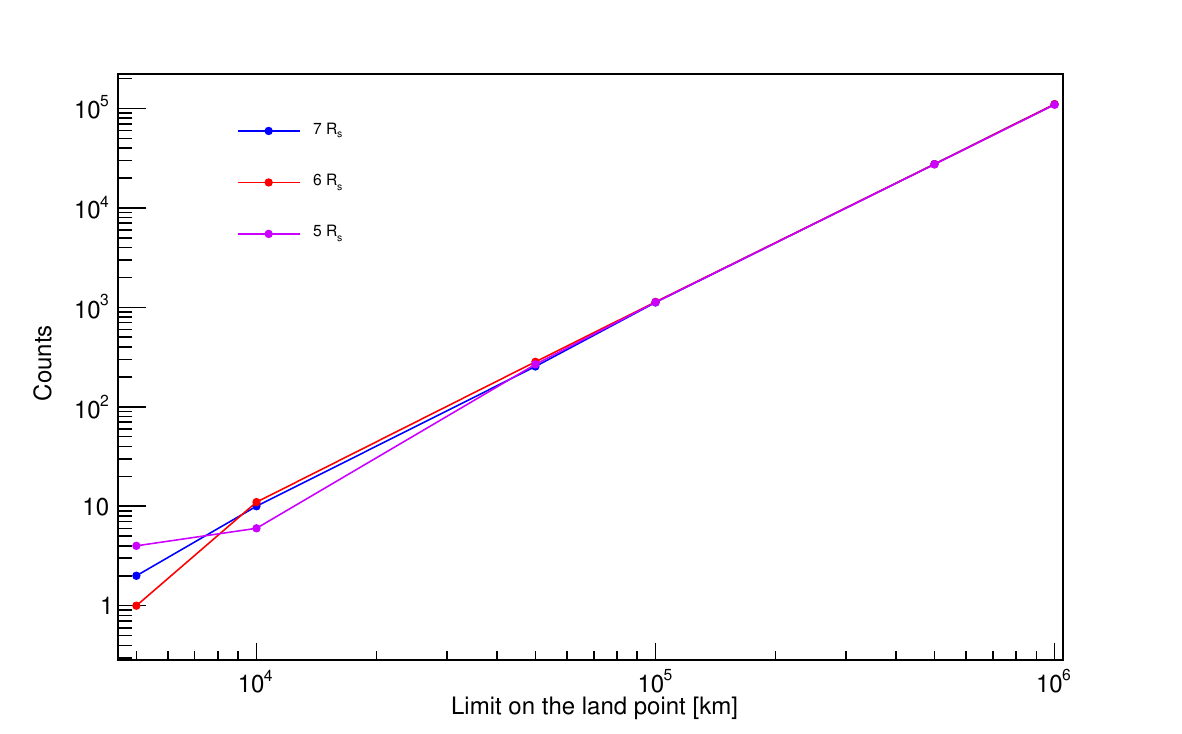}
        \caption{The graph of the number of simulations that fulfilled the land point limit in $10^{10}$ simulations, executed for each limit, as a function of the land point limit, for different start sphere radius: $5\ R_{\odot}$, $6\ R_{\odot}$ and $7\ R_{\odot}$}
        \label{fig: meet_limit}
    \end{figure}

\begin{figure}
        \includegraphics[width=10.5 cm]{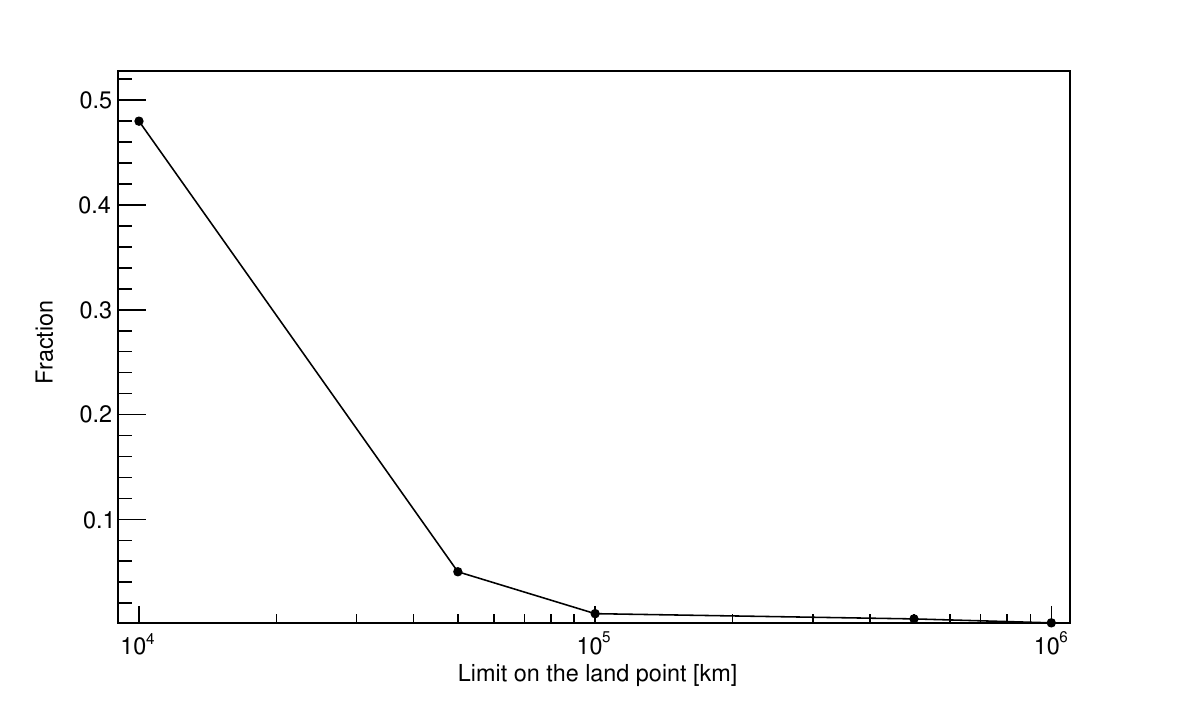}
        \caption{The graph of the fraction of simulations, for which at least some secondary photons reached the Earth's atmosphere, in all simulations that fulfilled the land point limit.}
        \label{fig: observed}
       \end{figure}

\begin{figure}
        \includegraphics[width=10.5 cm]{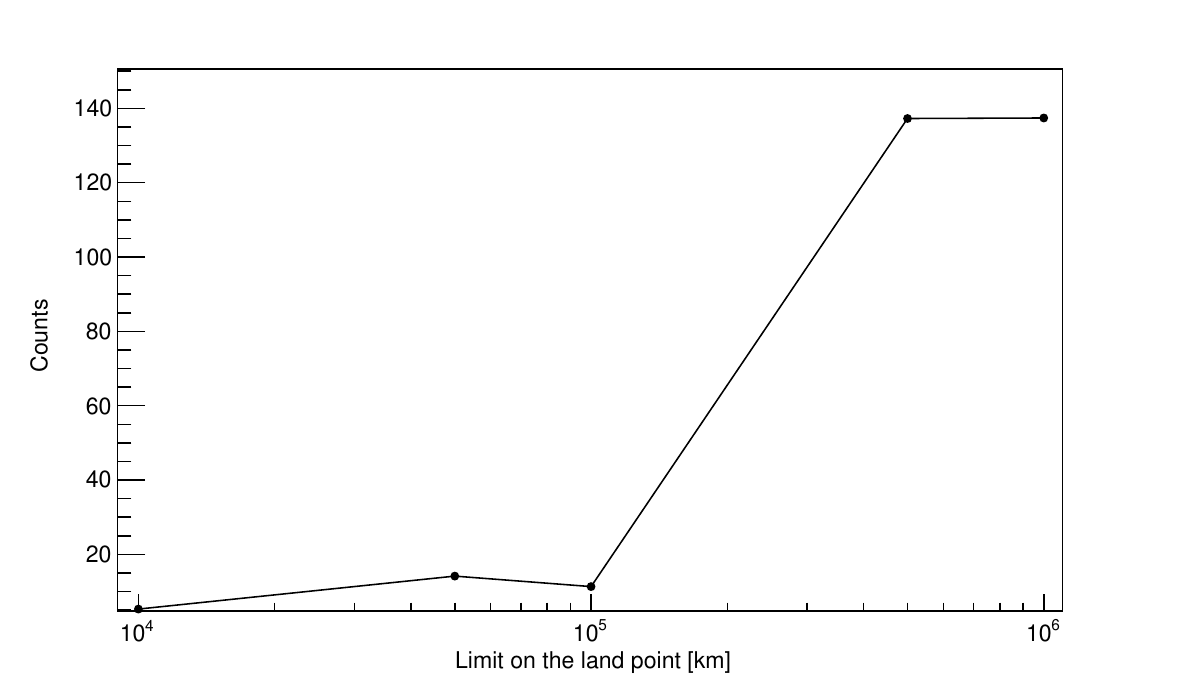}
        \caption{The calculated number of simulations giving observable effects in all executed simulations ($10^{10}$) as a function of the land point limit, on the basis of Figure~\ref{fig: meet_limit} and Figure~\ref{fig: observed}.}
        \label{fig: events_num_observed}
    \end{figure}

\begin{figure}
        \includegraphics[width=10.5 cm]{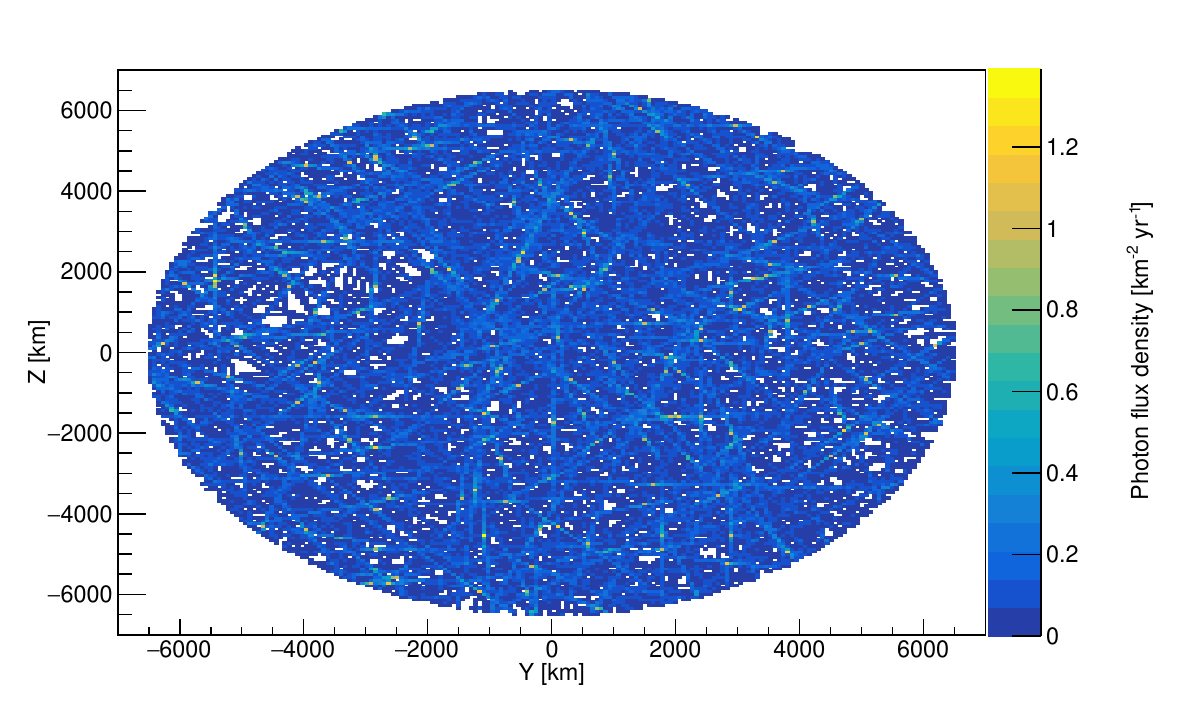}
        \caption{The cumulative spatial distribution of secondary photons at the top of the atmosphere, for the primary photons energy 100 EeV, based on 324 observable secondary photons cascades, which corresponds to one year of observation.}
        \label{fig: sps_last_cum}
    \end{figure}

\begin{figure}
        \includegraphics[width=10.5 cm]{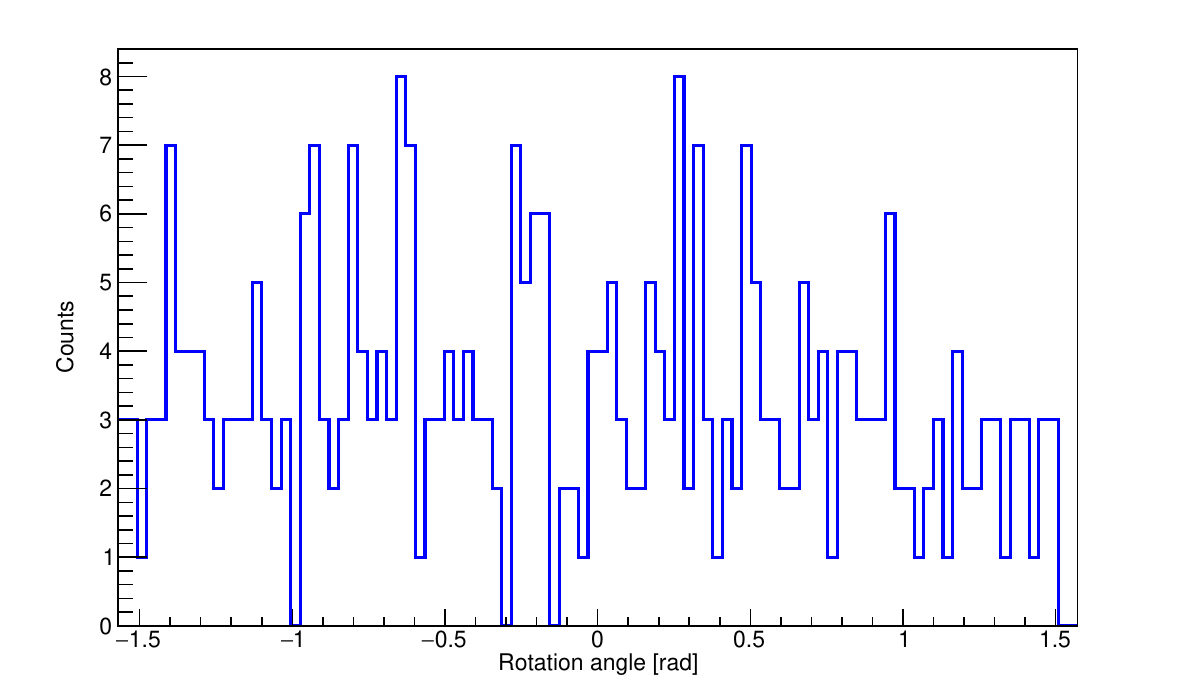}    
        \caption{The distribution of the orientation of CRE signatures at the top of the atmosphere, for the primary photons energy 100 EeV, based on 324 CRE signatures. Histograms are normalised and divided into 100 bins.}
        \label{fig: orient_cum}
    \end{figure}

    The implemented limit on the centre of the secondary photons spatial distribution and the start sphere radius - the simulation input parameter should be also adjusted for the optimization of the simulation.
    The limit on the primary photon land point has an influence on the ratio of observable on the Earth events to all simulations that fulfilled the condition. The number of simulations giving observable effects increases with the decrease in the distance from the $x$ axis to the centre of secondary photons signature. On the other hand, the limit obviously affects the fraction of simulations that fulfil the condition and in this way the computation time. The strengthening of the limit on the primary photon land point results in the decline in the ratio of simulation fulfilling the condition to all running simulations and considerably prolongs the calculation time. These relations are shown in Figure~\ref{fig: meet_limit} and Figure~\ref{fig: observed}.
    On the basis of the obtained data, we calculated the ratio of observable on the Earth events to all running simulations, which is illustrated in Figure~\ref{fig: events_num_observed}. It was observed that the fraction of detectable events is greater for the relaxation of the condition. However, the statistics for this range is based on few entries, so it may be burdened with high risk.
    Figure~\ref{fig: meet_limit} shows that the number of simulations fulfilling the condition depends weakly on the start sphere radius. Thus, the fraction of simulations observed on the Earth was calculated only for the start sphere radius 6 $R_{\odot}$ (Figure~\ref{fig: observed}).
    In order to consider the largest number of events observable on the Earth in a relatively short time, the land point limit was set at $10^4$ km.
    The number of CRE effects in all effects detectable on the Earth, in turn, is dependent on the start sphere radius. Due to the fact that the start sphere radius has an influence on primary photons' impact parameters, it also affects the probability of conversion. On the other hand, the decrease in the start sphere radius may also eliminate some cases that can give observable effects at the top of the atmosphere. Therefore the start sphere radius $6\ R_{\odot}$ was chosen as an optimum value for the simulation parameter. The resulting cumulative distribution of secondary photons at the top of the atmosphere is shown in Figure~\ref{fig: sps_last_cum}. 
    We enlarged the considered observable area to take into account the cross section of the Earth and the surrounding atmosphere. The distribution includes 324 CRE effect signatures, which corresponds to one year of observation taking into account present limits on the UHE photon flux \cite{Auger}.
    Observed line-like signatures are parts of large cascades with centres beyond the Earth's cross-section. They justified the consideration of the cumulative distribution of secondary photons. Due to the fact that cascades are produced by UHE photons in the vicinity of the Sun, many secondary photons have directions contained in the solid angle determined by the Sun. Therefore CRE effect cascades should distinguish themselves from the rest of the observations, since secondary photons may reach higher energies than photons emitted by the Sun. The anomaly in the very-high-energy photons flux was already observed (see \cite{HAWC} and \cite{Fermi-LAT}). However, the UHE photon cascading was not taken into account as a possible source to date. Therefore, it is significant to analyse the cumulative distribution of secondary photons with regard to unique features that could help to classify them. The possible anisotropy may be observed in the distribution of the orientation of signatures at the top of the atmosphere. The horizontal or nearly horizontal orientation could be more frequent. It would be also theoretically motivated since signatures oriented in this way are generated by UHE photons that have the heliocentric latitude around $0^{\circ}$ or $90^{\circ}$ at the conversion point \cite{Symmetry}. And for the heliocentric latitude $90^{\circ}$, the conversion occurs more often \cite{JCAP}. 
    In order to describe the orientation of signatures, the rotation angle parameter was introduced. It is given by the equation
    $$\alpha=\arctan\left(\frac{z_1-z_2}{y_1-y_2}\right),$$
    where points $(y_1,\ z_1)$ and $(y_2,\ z_2)$ are related to secondary photons that are the farthest, but still located in the observable area. The rotation angle is in the range of $(-\pi/2,\ \pi/2)$. Obtained distribution is shown in the Figure~\ref{fig: orient_cum}.
\linenumbers

\section{Discussion}
The previous study (\cite{JCAP, Symmetry}) has shown that the UHE photon interaction with the solar magnetic field may result in the extremely extended, line-like shaped cascade of secondary photons reaching the top of the Earth's atmosphere. Owing to the unique signature, the footprint of secondary photons at the top of the atmosphere should be possible to observe and classify. Due to the expected wide spatial span of these cascades, it should be possible to observe not only secondary photons produced by primary UHE photons aimed at the Earth, but also the tails of the secondary photons produced by primary UHE photons directed far from the Earth.
In this article we presented a successful method for modelling the interactions of the isotropic UHE photon flux with the solar magnetic field in order to obtain for the first time particle distributions containing secondary photons from both central and peripheral parts of the CRE cascades. The method is based on using the PRESHOWER program appropriately modified and optimized. We verified the functioning of the simulator and checked that obtained results are perfectly consistent with the previous analyses \cite{JCAP}. In order to investigate the cumulative distribution of secondary photons at the top of the atmosphere, we discussed the impact of simulation parameters on the modelled physical phenomena and found optimal values for these parameters.
The results obtained in this way indicate that secondary photons produced by the UHE photons not aimed at the Earth contribute significantly to the cumulative distribution of secondary photons at the top of the atmosphere and to the distributions of CRE footprint orientations which might serve as a valuable observational hint.
In relation to the methods described here it is important to note that recent observations (\cite{Fermi-LAT,HAWC}) demonstrate yet unexplained gamma-ray emission from the solar disk. In the light of the simulations of CRE resulting from the interactions of UHE photons nearby the Sun one might speculate that these observations may be a signature of the CRE effect which motivates a dedicated analysis in this direction. We conclude that it is justified and indispensable to simulate the flux of UHE photons cascading in the Sun's vicinity without neglecting the tails of the resultant CRE distributions, and that the presented method enable us to search for potentially observable signatures of the CRE effect, and hence offer a novel perspective for an indirect searches for UHE photons. We proposed that one of such signatures might be seen in a distribution of the spatial orientations of the CRE cascades at the top of the atmosphere. A quantitative evaluation of the appropriateness of this observable for the CRE and UHE photon searches would require a separate study which has already been planned.

\vspace{6pt} 



\authorcontributions{conceptualization, B.P., T.B., N.D., P.H., D.A.; investigation, B.P., T.B., N.D., P.H.; methodology, B.P., T.B., N.D., P.H.; software, B.P., N.D., P.H.; theoretical studies, B.P., T.B., D.A.; writing—original draft preparation, B.P., T.B., N.D., P.H.; writing–review and editing, B.P., T.B., N.D., O.S., S.S., P.H., M.P., T.W., J.S., P.K., K.S., M.R., M.N., J.M., T.S., \L{}.B., A.T., L.D., K.R.;
All authors have read and agreed to the published version of the manuscript.
}

\acknowledgments{
This research was supported in part by PLGrid Infrastructure and we warmly thank the staff at ACC Cyfronet AGH-UST for their always helpful supercomputing support.
}

\conflictsofinterest{
The authors declare no conflict of interest.
} 



\abbreviations{The following abbreviations are used in this manuscript:\\

\noindent 
\begin{tabular}{@{}ll}
UHE & Ultra-High Energy\\
CRE & Cosmic Ray Ensembles\\
EAS & Extensive Air Shower\\
CREDO & Cosmic-Ray Extremely Distributed Observatory
\end{tabular}}




\end{paracol}
\reftitle{References}


\externalbibliography{yes}
\bibliography{bibliography}

\begin{thebibliography}{999}

\bibitem[Homola and \textit{et al.} for~the CREDO~Collab(2020)]{Symmetry}
Homola, P.; \textit{et al.} for~the CREDO~Collab.
\newblock Cosmic-Ray Extremely Distributed Observatory.
\newblock {\em Symmetry} {\bf 2020}, {\em 12},~1835.
\newblock
  doi:{\changeurlcolor{black}\href{https://doi.org/10.3390/sym12111835}{\detokenize{10.3390/sym12111835}}}.

\bibitem[Greisen(1966)]{Greisen}
Greisen, K.
\newblock {End to the cosmic-ray spectrum?}
\newblock {\em Phys. Rev. Lett.} {\bf 1966}, {\em 1966},~748.

\bibitem[Zatsepin and Kuzmin(1966)]{Kuzmin}
Zatsepin, G.T.; Kuzmin, V.A.
\newblock {Upper limit of the spectrum of cosmic rays}.
\newblock {\em JETP Lett. 4, 78} {\bf 1966}, {\em 1966},~78.

\bibitem[Zyla and \textit{et al}(2020)]{PDG}
Zyla, P.; \textit{et al}.
\newblock {Review of Particle Physics}.
\newblock {\em PTEP} {\bf 2020}, {\em 2020},~083C01.
\newblock and 2021 update,
  doi:{\changeurlcolor{black}\href{https://doi.org/10.1093/ptep/ptaa104}{\detokenize{10.1093/ptep/ptaa104}}}.

\bibitem[Kuzmin and Rubakov(1998)]{SHDM_Kuzmin}
Kuzmin, V.; Rubakov, V.
\newblock Ultrahigh-energy cosmic rays: A window on postinflationary reheating
  epoch of the Universe?
\newblock {\em Physics of Atomic Nuclei - PHYS ATOM NUCL-ENGL TR} {\bf 1998},
  {\em 61},~1028--1030.

\bibitem[Birkel and Sarkar(1998)]{SHDM_Birkel}
Birkel, M.; Sarkar, S.
\newblock Extremely high energy cosmic rays from relic particle decays.
\newblock {\em Astroparticle Physics} {\bf 1998}, {\em 9},~297--309.
\newblock
  doi:{\changeurlcolor{black}\href{https://doi.org/https://doi.org/10.1016/S0927-6505(98)00028-0}{\detokenize{https://doi.org/10.1016/S0927-6505(98)00028-0}}}.

\bibitem[Frampton \em{et~al.}(1999)Frampton, Keszthelyi, and
  N~G]{SHDM_Frampton}
Frampton, P.H.; Keszthelyi, B.; N~G, Y.J.
\newblock Longevity and Highest-Energy Cosmic Rays.
\newblock {\em International Journal of Modern Physics D} {\bf 1999}, {\em
  08},~117--122,
  \href{http://xxx.lanl.gov/abs/https://doi.org/10.1142/S0218271899000109}{{\normalfont
  [https://doi.org/10.1142/S0218271899000109]}}.
\newblock
  doi:{\changeurlcolor{black}\href{https://doi.org/10.1142/S0218271899000109}{\detokenize{10.1142/S0218271899000109}}}.

\bibitem[{Blasi} \em{et~al.}(2002){Blasi}, {Dick}, and {Kolb}]{SHDM_Blasi}
{Blasi}, P.; {Dick}, R.; {Kolb}, E.W.
\newblock {Ultra-high energy cosmic rays from annihilation of superheavy dark
  matter}.
\newblock {\em Astroparticle Physics} {\bf 2002}, {\em 18},~57--66,
  \href{http://xxx.lanl.gov/abs/astro-ph/0105232}{{\normalfont
  [arXiv:astro-ph/astro-ph/0105232]}}.
\newblock
  doi:{\changeurlcolor{black}\href{https://doi.org/10.1016/S0927-6505(02)00113-5}{\detokenize{10.1016/S0927-6505(02)00113-5}}}.

\bibitem[det()]{detector}
CREDO Detector.
\newblock \url{https://credo.science/#/detector/tutorial}.

\bibitem[Bibrzycki and \textit{et al.} for~the
  CREDO~Collab(2020)]{detector_paper}
Bibrzycki, L.; \textit{et al.} for~the CREDO~Collab.
\newblock Towards A Global Cosmic Ray Sensor Network: CREDO Detector as the
  First Open-Source Mobile Application Enabling Detection of Penetrating
  Radiation.
\newblock {\em Symmetry} {\bf 2020}, {\em 12}.
\newblock
  doi:{\changeurlcolor{black}\href{https://doi.org/10.3390/sym12111802}{\detokenize{10.3390/sym12111802}}}.

\bibitem[Dhital \em{et~al.}(2022)Dhital, Homola, Alvarez-Castillo, G\'ora,
  Wilczy\'nski, Almeida~Cheminant, Poncyljusz, M\k{e}drala, Opi\l{}a, and
  \textit{et al.} for~the CREDO~Collab]{JCAP}
Dhital, N.; Homola, P.; Alvarez-Castillo, D.; G\'ora, D.; Wilczy\'nski, H.;
  Almeida~Cheminant, K.; Poncyljusz, B.; M\k{e}drala, J.; Opi\l{}a, G.;
  \textit{et al.} for~the CREDO~Collab.
\newblock Cosmic ray ensembles as signatures of ultra-high energy photons
  interacting with the solar magnetic field.
\newblock {\em Journal of Cosmology and Astroparticle Physics} {\bf 2022}, {\em
  2022},~038.
\newblock
  doi:{\changeurlcolor{black}\href{https://doi.org/10.1088/1475-7516/2022/03/038}{\detokenize{10.1088/1475-7516/2022/03/038}}}.

\bibitem[Homola \em{et~al.}(2005)Homola, G\'ora, Heck, Klages, P\k{e}kala,
  Risse, Wilczy\'nska, and Wilczy\'nski]{preshower2005}
Homola, P.; G\'ora, D.; Heck, D.; Klages, H.; P\k{e}kala, J.; Risse, M.;
  Wilczy\'nska, B.; Wilczy\'nski, H.
\newblock Simulation of ultra-high energy photon propagation in the geomagnetic
  field.
\newblock {\em Computer Physics Communications} {\bf 2005}, {\em 173},~71--90.
\newblock
  doi:{\changeurlcolor{black}\href{https://doi.org/https://doi.org/10.1016/j.cpc.2005.07.001}{\detokenize{https://doi.org/10.1016/j.cpc.2005.07.001}}}.

\bibitem[Homola \em{et~al.}(2013)Homola, Engel, Pysz, and
  Wilczyński]{PRESHOWER}
Homola, P.; Engel, R.; Pysz, A.; Wilczyński, H.
\newblock Simulation of ultra-high energy photon propagation with PRESHOWER
  2.0.
\newblock {\em Computer Physics Communications} {\bf 2013}, {\em 184}.
\newblock
  doi:{\changeurlcolor{black}\href{https://doi.org/10.1016/j.cpc.2013.01.015}{\detokenize{10.1016/j.cpc.2013.01.015}}}.

\bibitem[Rautenberg(2019)]{Auger}
Rautenberg, J.
\newblock Limits on ultra-high energy photons with the Pierre Auger
  Observatory.
\newblock  2019, p. 398.
\newblock
  doi:{\changeurlcolor{black}\href{https://doi.org/10.22323/1.358.0398}{\detokenize{10.22323/1.358.0398}}}.

\bibitem[Erber(1966)]{Erber}
Erber, T.
\newblock High-Energy Electromagnetic Conversion Processes in Intense Magnetic
  Fields.
\newblock {\em Reviews of Modern Physics - REV MOD PHYS} {\bf 1966}, {\em
  38},~626--659.
\newblock
  doi:{\changeurlcolor{black}\href{https://doi.org/10.1103/RevModPhys.38.626}{\detokenize{10.1103/RevModPhys.38.626}}}.

\bibitem[Daugherty and Harding(1983)]{Daughtery}
Daugherty, J.; Harding, A.
\newblock Pair production in superstrong magnetic fields.
\newblock {\em ApJ} {\bf 1983}, {\em 273}.
\newblock
  doi:{\changeurlcolor{black}\href{https://doi.org/10.1086/161411}{\detokenize{10.1086/161411}}}.

\bibitem[Sokolov \em{et~al.}(2022)Sokolov, Ternov, Kilmister, and
  Chomet]{Sokolov}
Sokolov, A.; Ternov, I.; Kilmister, C.; Chomet, S.
\newblock Radiation from relativistic electrons / A. A. Sokolov, I. M. Ternov.
\newblock {\em SERBIULA (sistema Librum 2.0)} {\bf 2022}.

\bibitem[Wiedemann(2003)]{Wiedemann}
Wiedemann, H.
\newblock {\em Synchrotron Radiation}; Advanced Texts in Physics, Springer,
  2003.

\bibitem[Albert and \textit{et al.} for~the Auger~Collab(2018)]{HAWC}
Albert, A.; \textit{et al.} for~the Auger~Collab.
\newblock First HAWC observations of the Sun constrain steady TeV gamma-ray
  emission.
\newblock {\em Physical Review D} {\bf 2018}, {\em 98}.
\newblock
  doi:{\changeurlcolor{black}\href{https://doi.org/10.1103/PhysRevD.98.123011}{\detokenize{10.1103/PhysRevD.98.123011}}}.

\bibitem[Rainó \em{et~al.}(2017)Rainó, Giglietto, Moskalenko, Orlando, and
  Strong]{Fermi-LAT}
Rainó, S.; Giglietto, N.; Moskalenko, I.; Orlando, E.; Strong, A.
\newblock Fermi Large Area Telescope Observations of the gamma-ray emission
  from the Quiescent Sun.
\newblock {\em Nuclear and Particle Physics Proceedings} {\bf 2017}, {\em
  291-293},~36--39.
\newblock
  doi:{\changeurlcolor{black}\href{https://doi.org/10.1016/j.nuclphysbps.2017.06.008}{\detokenize{10.1016/j.nuclphysbps.2017.06.008}}}.

\end{thebibliography}

\end{document}